# O-acetylated sugars in the gas phase: stability, migration, positional isomers and conformation †

Oznur Yeni,[a+] Amira Gharbi,[a+] Stéphane Chambert,[b] Jean Rouillon,[b] Abdul-Rahman Allouche,[a] Baptiste Schindler[a] and Isabelle Compagnon*[a]

O-acetylations are functional modifications which can be found on different hydroxyl groups of glycans and which contribute to the fine tuning of their biological activity. Localizing the acetyl modifications is notoriously challenging in glycoanalysis, in particular because of their mobility: loss or migration of the acetyl group may occur through the analytical workflow. Whereas migration conditions in the condensed phase have been rationalized, little is known about the suitability of Mass Spectrometry to retain and resolve the structure of O-acetylated glycan isomers. Here we use the resolving power of infrared ion spectroscopy in combination with ab initio calculation to assess the structure of O-acetylated monosaccharide ions in the gaseous environment of a mass analyzer. *N*-acetyl glucosamines were synthetized with an O-acetyl group in positions 3 or 6, respectively. The protonated ions produced by electrospray ionization were observed by mass spectrometry and their vibrational fingerprints were recorded in the 3 μm range by IRMPD spectroscopy (InfraRed Multiple Photon Dissociation). Experimentally, the isomers show distinctive IR fingerprints. Additionally, ab initio calculations confirm the position of the O-acetylation and resolve their gas phase conformation. These findings demonstrate that the position of O-acetyl groups is retained through the transfer from solution to the gas phase, and can be identified by IRMPD spectroscopy.

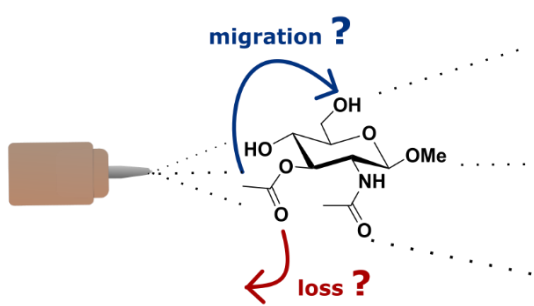

## Introduction

Glycans play a key role in various physiological and pathological processes including protein folding, cell-pathogen interaction, cancer metastasis and immune response.[1]

Therefore, thorough analysis of glycan structure is needed to better understand the impact of their structural variations on their biological effect. However, it is a challenge to characterize all sequences of glycans and polysaccharides because of their structural diversity: there are different types of monosaccharides that can be connected with different linkage, resulting in a tremendous number of possible combinations.[2]

Added to that, saccharides can be substituted with non-sugar constituents like O-acetyl, which also contribute to their biological activity. For example, it was reported that the degree of O-acetylation can alter antigenicity and immunogenicity for some bacterial species.[3] Such is the case of *Salmonella typhi* : O-acetylation is necessary on Vi capsular polysaccharide to induce antibody synthesis.[4] On the other hand, a decrease of O-acetylated sialic acids was reported in colorectal carcinoma.[5]

Therefore, the determination of precise carbohydrate sequence supports the optimization of vaccine design and the discovery of cancer biomarkers.[6,7]

Acetyl groups can be positioned on the different hydroxyl groups of the sugar ring. NMR studies have shown that *Shigella flexneri* lipopolysaccharide presents different O-acetyl localisation: at O3 of α-Rha*p* and O6 of β-GlcNAc, with different degree.[8] Other studies on *Neisseria* have shown that on terminal GlcNAc of its lipooligosaccharide, O-acetylation is located at the O3 position.[9] Many other examples of bacteria with different O-acetyl patterns on different residues have been reported.[10–12]

Contrastingly, the impact of the acetyl position on its biological environment was not as thoroughly investigated. Fusco *et al.* have shown that group C meningococcal polysaccharide conjugate vaccine acetylated on C-7 of sialic acid provided a better bacterial inhibition than C-8. The authors suggested that OAc group can influence the conformational preference depending on its proximity to the glycosidic bond.[13] It was recently reported that the main acetylated sialic acid in breast cancer cells is substituted at position C9.[14] The regiochemistry of O-acetylation is a minute detail, but can be critical. Hence, developing tools to identify it will lead to a better understanding of its biological activity.

However, a difficulty to be considered for such study is the potential loss or migration of acetyl group at some point of the analytical workflow, which can lead to error on sequence determination. It is established that some conditions in solution can induce acetyl migration. For example, exposure to alkaline condition results in migration of acetyl group from C7 to C9 for sialic acid[15] and from C3 to C6 for some hexoses.[16] The fact that a number of previously identified acetyl positions on capsular polysaccharides were revised[17,18] is indicative of the complexity of this issue. Therefore, it is important to take precautions throughout the analytical workflow.

In case of MS analysis, the necessity to transfer the ions into gas phase opens additional questions. Gas-phase rearrangement reactions were reported in proteomics[19] and glycomics.[20] It may concern an entire hexose like fucose[21] or a chemical substituent like sulphate.[22] In these studies, it has been proposed that collisional induced dissociation (CID) can generate ion rearrangement. Recently, Pagel et al. also suggested that migration of fucose might occur prior to CID.[23]

By adding a spectroscopic dimension to mass spectrometry, IRMPD spectroscopy (InfraRed Multiple Photon Dissociation) or other variants of IR ion spectroscopy can resolve isobaric and isomeric ions because they generally feature distinctive vibrational patterns. After building a library of reference vibrational fingerprints for a limited number of glycan standards, it is possible to elucidate the monosaccharide content; regiochemistry and stereochemistry of the glycosidic bonds of an unknown oligosaccharide.[24,25]

This method has been recently applied in the case of positional isomers of sulphated saccharides. In the IR range offered by a table-top tuneable laser systems (around 3μm), sulphate group vibrations are not directly observed. Yet, this substituent affects the spatial disposition of diagnostic X-H vibrations, including N-H, O-H and C-H stretches. It was thus possible to identify sulphate position by investigating NH, OH and CH vibrational patterns.[26] To the best of our knowledge, gas phase ion spectroscopy has not yet been applied to the case of positional isomers of acetylated glycans.

Here we propose an exploration of the behaviour of O-acetylated monosaccharides models in the gaseous environment of a mass analyzer: is the position of the acetyl group retained through the transfer from solution to the gas phase? Is it possible to resolve positional isomer? How does the acetyl group affect the conformation of the glycan? To that end, we use a combination of IRMPD spectroscopy and *ab initio* calculations to characterize the structures of synthetic standards of glucosamine with acetyl groups in various positions.

## Methods

### Synthesis

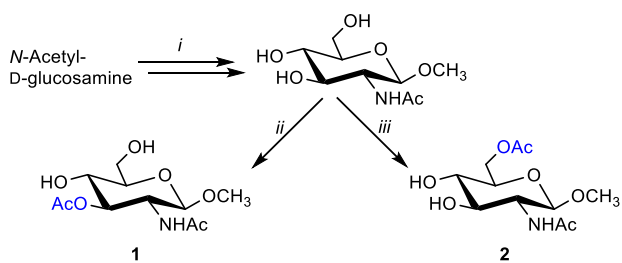

**Scheme 1.** Synthesis of the two *O*-monoacetylated *N*-acetyl-D-glucosamines β-GlcNAcOMe3OAc (**1**) and β-GlcNAcOMe6OAc (**2**). Reagents and conditions: *i*. Acetone/FeCl₃, reflux, 20 min, then methanol/p-TsOH (according to Furneaux et al. and Cai et al.)[27,28]; *ii*. Dimethoxytrityl chloride/pyridine, then AcCl and dichloroacetic acid (62%); *iii*. PPh₃, diisopropyl azodicarboxylate (DIAD), AcOH, toluene/DMF (39%).

*N*-acetyl-D-glucosamine was transformed into the key β-*O*-methyl glycoside intermediate according to a procedure developed by Bundle et al. (Scheme 1).[28] This synthesis involved first the formation of a furanosyl oxazoline[27] which was then opened under acid catalysis to give the desired β-*O*-methyl glycoside. To obtain the desired regioselective *O*-acetylations, two pathways were followed. For the synthesis of the 3-*O*-acetylated derivative **1**, a dimethoxytrityl group was first installed on position 6, followed by acetylation on position 3. Even with an excess of acetyl chloride, no further acetylation was observed in position 4 (followed by TLC). Therefore, a direct deprotection was set up with a 2% dichloroacetic solution, in a short time to avoid migration of acetyl and the 3-*O*-acetylated compound **1** was obtained with 62% yield.

The synthesis of the 6-*O*-acetylated derivative **2** takes advantages of the selective reactivity of the primary alcohol function in Mitsunobu conditions. Using diisopropyl azodicarboxylate (DIAD) and triphenylphosphine gave the selective formation of the 6-*O*-acetyl ester **2** in 39% yield.

Knowing the assignment of the characteristic signals of β-GlcNAcOMe,[28] the selectivity of the monoacetylation of compounds **1** and **2** is confirmed by comparing the different ¹H NMR spectra (figure S5): i/ the protons associated with the acetyl bearing positions are drastically deshielded (δ(H3) going from 3.55 ppm for the non-acetylated β-GlcNAcOMe and 4.96 ppm for **1**; δ(H6a) and δ(H6b) going from 3.94 ppm and 3.77 ppm for the non-acetylated β-GlcNAcOMe and 4.41 ppm and 4.22 ppm respectively for **2**). ii/ the appearance of a single additional signal in the acetyl region (2.04 ppm for **1** and 2.07 ppm for **2**).

### Experimental detail

Synthetic monosaccharides **1** and **2** were prepared in methanol 100% to avoid migration in solution at a concentration of 40 μmol/L.

The samples were introduced in a modified commercial linear ion trap LTQ XL equipped with electrospray ion source (ESI) and enabled with IRMPD spectroscopy, that was previously described.[29]

The samples were infused and ionized by ESI in positive mode to yield protonated ions. The ions of interest were isolated in an ion trap that was drilled and equipped with an IR transparent window, allowing for irradiation of the ion cloud by a tunable YAG-pumped OPO/OPA IR laser system. If laser excitation corresponds to a vibrational mode of the ion, its internal energy increases and then relaxes by photofragmentation. The photofragmentation yield is calculated according to the formula $-\log(I_{parent}/(I_{parent} + I_{fragment}))$ with $I_{parent}$ the intensity of the parent ion and $I_{fragment}$ is the summed intensity of fragments detected after laser irradiation at each wavelength between 2700 and 3700 cm⁻¹. On the IRMPD spectra shown thereafter, data are averaged 3 times and a line of 5 points smooth FFT is added to guide eye.

### Computational detail

Molecular Dynamics was used to explore the potential energy surface of the protonated ions with the PM7 potential in OpenMopac.[30] For each starting structure, 1 trajectory of 10 ps at 4000 K was ran and yielded 2000 geometries. After optimization with PM7 and elimination of the identical conformers, around 1000 geometries were obtained. The geometries were firstly optimized with B3LYP[31–33]/6-31+G*, which resulted in further reduction of the number of stable geometries (around 400 conformations). The conformers are sorted by electronic energy (the zero-point energy and temperature effect are not included). Then the 50 lowest energy conformers were optimized and harmonic frequencies computed with CAM-B3LYP[34]/6-311++G(2df,2pd).[35–37] An empirical scaling factor of 0.947[38] was used for comparison with the experimental IRMPD spectra. The mode analysis was performed using Gabedit,[39] Gaussian09[40] was used for DFT calculations.

## Results and discussion

### MS/MS and IRMPD signatures

Protonated monosaccharides **1** and **2** were detected by mass spectrometry at m/z 278.

Fig.1a shows the MS/MS spectra obtained by collision induced dissociation (CID) for **1** (green) and **2** (orange). For each isomer, the main fragment (m/z 246) corresponds to the loss of the O-methyl group. For the compound acetylated on O3 (**1**), a minor fragment was observed at m/z=186: it corresponds to the loss of both O-methyl and O-acetyl groups. Potentially isomer-diagnostic cross ring fragments were not observed in MS/MS using CID (nor in MS³, data shown in Fig.S6). The MS/MS data clearly show that the O-acetyl group does not easily fragment, therefore it is stable once in the gaseous environment.

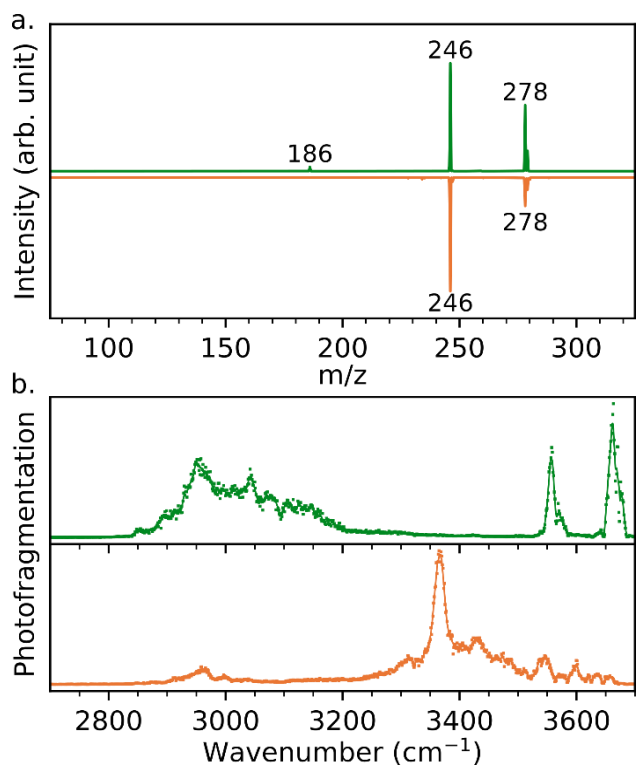

Figure 1 a) MS/MS spectra of protonated monosaccharides **1** (green) and **2** (orange); b) IRMPD spectra of protonated monosaccharides **1** (green) and **2** (orange)

The IRMPD spectra of protonated standards are displayed in Fig 1b. For monosaccharide **1**, the IRMPD spectrum shows a broad region of vibrational activity from 2850 to 3200 cm$^{-1}$, which corresponds to an unresolved ensemble of NH and CH stretching modes. We also observe a doublet of resolved features in the OH stretching range at 3558 and 3663 cm$^{-1}$. For monosaccharide **2**, IR activity below 3200 cm$^{-1}$ is weak, with two features at 2960 and 3000 cm$^{-1}$ corresponding to CH stretching modes. IR absorption is mainly observed above 3200 cm$^{-1}$ with an unresolved pattern featuring an intense peak at 3365 cm$^{-1}$ and several minor peaks.

The two vibrational fingerprints are distinctive, which implies that the two protonated ions feature distinctive structures. This suggests that they retain their individual structures, thus acetyl migration has not occurred during the transfer from solution to the gas phase. Yet, molecular rearrangements might occur and yield distinctive products with individual IRMPD spectra. It is thus essential to correlate the IRMPD fingerprint to the exact molecular structure, in order to draw reliable conclusions concerning possible gas phase rearrangements.

While the interpretation of IRMPD spectra generally requires quantum chemistry tools, the spectrum of **1** shows some features which are relatively intuitive and allow for a hand-waving interpretation: with a hexose methylated at O1, N-acetylated at O2 and O-acetylated at O3, we are left with 2 OH groups (expectedly O4H and O6H). The two resolved bands at 3560 and 3660 cm$^{-1}$ are consistent with a structure featuring two OH groups. Based on prior experience,[41,42] one can tentatively assign the band at 3660 cm$^{-1}$ to the O6H vibration. The presence of a free O6H group is consistent with **1**. It further suggests that this structure was retained in the gas phase. The IRMPD spectrum of **2** does not show such a simple pattern of OH bands, thus empirical interpretation cannot be proposed.

Our results show individual IRMPD fingerprints for the two O-acetylated models, demonstrating that the two standards retain distinctive structures in gas phase and suggesting that migration does not occur through the ionization, transfer and trapping. While the fingerprint of compound **1** features a typical free O6H band (which is consistent with GlcNAcOMe acetylated on O3), the IR fingerprint of **2** does not support hand-waving interpretation. Therefore, these vibrational fingerprints must be interpreted in terms of molecular structure with the support of ab initio calculations, in order to fully validate the absence of acetyl migration.

**Identification with quantum chemistry calculations**

Quantum chemistry calculations in combination with IRMPD spectroscopy are generally used to resolve the conformation of gas phase ions.[43] Here, we primarily aim at validating the covalent structure (positional isomers) of our ions before describing their conformation.

Barnes *et al.* have previously reported the gas phase conformations of protonated *N*-acetylhexosamines.[42] GlcNAc was found to protonate on the carbonyl group of the *N*-acetyl group, and to adopt a $^4C_1$ conformation at low energy. Here, the presence of a second acetyl group may affect both the proton localization and the conformation. Therefore, for each positional isomer a pair of starting structures with a proton located on the *N*-acetyl function and on the *O*-acetyl function, respectively, was submitted to conformational exploration. When structures resulting from the conformational exploration were optimized, most starting structures featuring the proton on the OAc group saw a quick transfer of the proton towards the NAc group. Only higher energy conformations (>40 kj/mol) retained a proton on the OAc group. The later are shown in SI (Fig.S7), and only the former are discussed below.

Selected low energy structures of β-GlcNAcOMe acetylated on O3 or O6 are shown in figure 2. We can notice that similarly to GlcNAc without O-acetylation,[42] the low-energy structures adopt $^4C_1$ chair conformation, regardless of the acetyl position. For β-GlcNAcOMe3OAc (**1**), the lowest energy conformer $^4C_1$-3OAc is stabilized by three main hydrogen bonds (not shown in Fig.2 for clarity, see SI). Knowing that the frequency of H-bonded donor group is redshifted, the proximity between NH and 3OAc group permitting C=O---HN bond can explain the lower NH mode observed at 3050 cm$^{-1}$. Similarly, the *N*-acetyl OH is strongly bound to the OMe group, resulting in a redshifted mode at 2850 cm$^{-1}$. Note that experimentally, H-bonded modes have a lower intensity and may not be observable.[44] The H-bond between the two OH groups O4H---O6 explains that the O4H feature has a lower wavenumber (3555 cm$^{-1}$) than O6H (3670 cm$^{-1}$). In contrast, the rotation of O4H and O6H on the second lowest energy conformer $^4C_1$-3OAc' impedes the hydrogen bond, so the two OH are free and correspond to higher frequencies around 3650 cm$^{-1}$. On this conformation, we still observe C=O---HN hydrogen bond but the rotation of OAc moves the oxygen atom away from the NH group, which explains that NH frequency is a little higher for this conformation (3110 cm$^{-1}$).

We also observe skew conformations for **1** at higher energies. The skew conformation with lowest energy $^1S_5$-3OAc is stabilized by O4H---O6 hydrogen bond and its OH features are similar to those of $^4C_1$-3OAc.

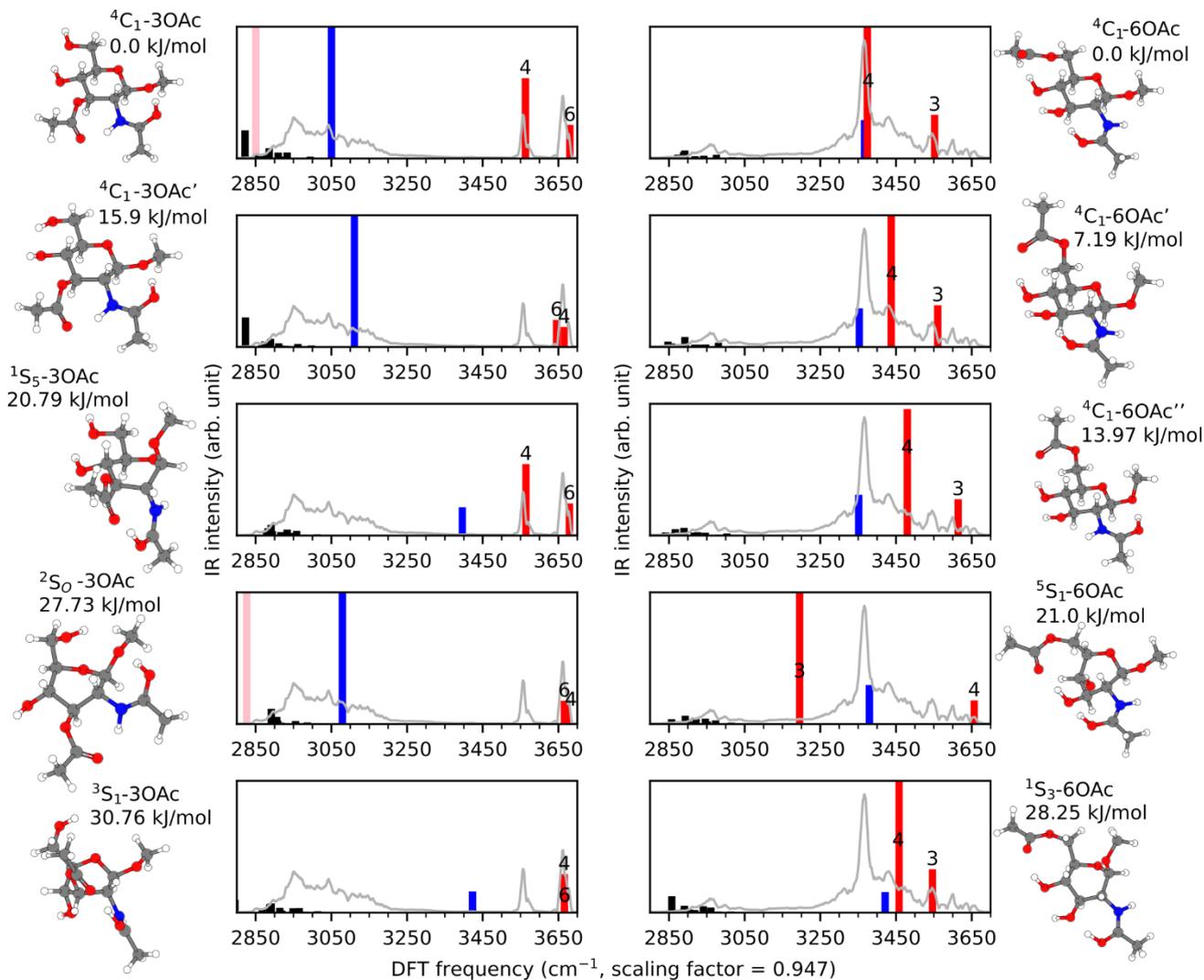

Figure 2 Experimental spectrum (grey line) compared to computational IR spectra (bars) of 5 conformations of protonated β-GlcNAcOMe3OAc (left) and 6OAc (right). OH stretches are colored in red (with label indicating its position on the ring), NH in blue, CH in black and *N*-acetyl OH in pink. For each conformer, the structure is represented and its conformation and relative energy calculated with CAM-B3LYP are indicated.

However, NH rotation hinders the interaction with OAc, which results in a higher frequency around 3400 cm$^{-1}$. Other skew conformation with free OH and free NH ($^2S_0$-3OAc) or H-bonded NH ($^3S_1$-3OAc) are observed at higher energy. Overall, the comparison between the simulated spectra and the experimental spectrum (grey line) shows an excellent match with the lowest energy conformation $^4C_1$-3OAc: the doublet of experimental features at 3560 and 3660 cm$^{-1}$ matches the calculated frequencies for O4H and OH6, respectively; and the large band from 2900 to 3200 cm$^{-1}$ can be explained by unresolved CH and strongly H-bonded NH stretching vibrations. The presence of $^4C_1$-3OAc' and $^2S_0$-3OAc cannot be excluded based on their vibrational patterns, but their higher relative energy suggests that they would be present in minor amount, if any. $^1S_5$-3OAc and $^3S_1$-3OAc are excluded based on their NH frequency.

For β-GlcNAcOMe6OAc (**2**), the lowest energy structure is a $^4C_1$ chair conformation with a full hydrogen bonding network O1---HNCCH$_3$OH---O3H---O4H---OCH$_3$O6. The second and third lowest energy structures consist of rotation of OAc and NAc groups. The three structures share a similar IR fingerprint with a free NH band matching the experimental feature at 3365 cm$^{-1}$; a H-bonded O4H mode ranging from 3300 to 3475 cm$^{-1}$, which accounts for the broad experimental feature; and a free or weakly bond O3H mode which accounts for the smaller experimental features above 3500 cm$^{-1}$. Here again skew structures were observed at higher energy and discarded.

The comparison with quantum chemistry simulations confirms that experimental spectra correspond respectively to protonated **1** and **2**. The position of OAc modifies the conformation and hydrogen network, so it impacts NH, OH and CH stretches. Thus, indirect identification may be possible in the 3μm range. They have a diagnostic fingerprint: the two OH features at 3560 and 3660 cm$^{-1}$ for **1** and the large band from 3300 to 3475 cm$^{-1}$ for **2**.

## Conclusions

In this work, we addressed the stability of the fragile o-acetylation of sugars through a MS analytical workflow. To that end, we used a combination of IRMPD and *ab initio* calculations to study the gas phase structure of two isomers: β-GlcNAcOMe acetylated on O3 or O6.

First of all, MS/MS spectra have shown that fragmented ions largely retain their acetyl group through collisional induced dissociation, which indicates that O-acetyl groups are remarkably stable in a gaseous environment.

Secondly, IRMPD spectra featuring a distinctive fingerprint for each isomer were obtained, suggesting that **1** and **2** retained their structure. The match between the IRMPD spectra and the calculated frequencies of the lowest energy conformers for each isomer further confirmed that the position of acetyl in solution is retained through the transfer to the gas phase by electrospray ionisation. The site of protonation and the gas phase conformation were also resolved.

Finally, our results demonstrate that intact O-acetylated sugars can be transferred from solution to the gas phase by electrospray ionization as a starting point of an MS analytical workflow, without loss or migration of the acetyl group. Additionally, we report that positional isomers feature diagnostic vibrational fingerprints and can be identified using IRMPD spectroscopy. We expect that the spectroscopic fingerprints obtained on standards can be used as references to determine acetyl position on GlcNAc from bacteria or cancer cells. A similar approach could be adopted to explore the acetylation of other compounds, such as sialic acid, which is involved in many biological processes.

## Author contributions

O.Y.: Investigation, Visualization, Writing - Original Draft, Writing - Review & Editing, Visualization. A.G.: Investigation. S.C. and J.R: Resources. A-R.A.: Resources. B.S.: Visualization. I.C.: Conceptualization, Writing - Original Draft, Writing - Review & Editing, Supervision, Project administration, Funding acquisition.

+ O.Y. and A.G. contributed equally to this work.

ORCID:
Oznur Yeni: 0000-0001-6171-3096
Amira Gharbi : 0000-0002-7833-6204
Abdul-Rahman Allouche : 0000-0003-0725-4057
Baptiste Schindler : 0000-0002-7376-4154
Isabelle Compagnon : 0000-0003-2994-3961

## Conflicts of interest

There are no conflicts to declare.

## Acknowledgements

The authors thank the Agence Nationale de la Recherche (ANR) for financially supporting ALGAIMS (ANR-18-CE29-0006-02) and GEPHIR (237038-UMR5306) projects. This work was granted access to the HPC resources of the "Centre de calcul CC-IN2P3" at Villeurbanne, France.

## Notes and references